\documentclass[preprint,12pt]{elsarticle}

\usepackage{amssymb}
\usepackage{amsmath}
\usepackage[utf8]{inputenc}
\usepackage[T1]{fontenc} 
\usepackage{commath}

\usepackage{setspace}
\usepackage{graphicx}
\usepackage{siunitx}
\usepackage[nomarkers,figuresonly]{endfloat}

\journal{Scripta Materialia}

\begin{document}

\begin{frontmatter}

\title{Investigation of the orientation relationship between nano-sized G-phase precipitates and austenite with scanning nano-beam electron diffraction using a pixelated detector}


\author[p:mpie]{Niels Cautaerts}
\author[p:gren]{Edgar F. Rauch}
\author[p:mpie]{Jiwon Jeong}
\author[p:mpie]{Gerhard Dehm}
\author[p:mpie]{Christian H. Liebscher}

\address[p:mpie]{Advanced Transmission Electron Microscopy, Structure and Nano-/ Micromechanics of Materials, Max-Planck-Institut für Eisenforschung GmbH, Max-Planck-Straße 1, 40237, Düsseldorf, Germany}
\address[p:gren]{Génie Physique et Mécanique des Matériaux, Science et Ingénierie des Matériaux et Procédés, Grenoble INP, 1130 rue de la Piscine, FR-38402, Saint Martin d'Hères Cedex, France}


\begin{abstract}
Scanning nano-beam electron diffraction with a pixelated detector was employed to investigate the orientation relationship of nanometer sized, irradiation induced G-phase (M$_6$Ni$_{16}$Si$_{7}$) precipitates in an austenite matrix.
Using this detector, the faint diffraction spots originating from the small G-phase particles could be resolved simultaneously as the intense matrix reflections.
The diffraction patterns were analyzed using a two-stage template matching scheme, whereby the matrix is indexed first and the precipitates are indexed second after subtraction of the matrix contribution to the diffraction patterns.
The results show that G-phase forms with orientation relationships relative to austenite that are characteristic of face-centered cubic (FCC) to body-centered cubic (BCC) transformations.
This work demonstrates that nano-beam electron diffraction with a pixelated detector is a promising technique to investigate orientation relationships of nano-sized precipitates with complex crystal structures in other material systems with relative ease.
\end{abstract}


\end{frontmatter}

\newpage



\doublespacing{}
G-phase is an intermetallic silicide with a complex cubic crystal structure (space group Fm$\bar{3}$m, lattice parameter around 1.1 nm)~\cite{yan2009crystal} and a stoichiometry represented by the formula M$_6$X$_{16}$Si$_{7}$ with M and X representing multiple possible metals.
G-phase can be precipitated in various steels, with M=Ti, Mn, Cr and X=Ni, Fe being the most common varieties.
In duplex steels subjected to ageing heat treatment, the phase precipitates in the ferrite phase with the cube-on-cube orientation relationship~\cite{auger1990atom, mateo1997characterization, spiegel1963ternary}.
Large G-phase domains on grain boundaries are associated with embrittlement~\cite{Shuro2012, shiao1994aging}, but precipitation of G-phase on dislocations can improve strength and creep resistance~\cite{sun2017g}.
Hence G-phase is of high technological importance in ferritic and duplex stainless steels.

In austenitic stainless steels, G-phase (M$_{6}$Ni$_{16}$Si$_7$) does not form under any ageing conditions, but it readily forms under irradiation over a wide temperature range ($\sim$300--700$^\circ$C)~\cite{Lee1980, Boothby1987, Lee1983}.
This is detrimental in nuclear applications, where the formation of G-phase in austenitic steels is associated with void swelling~\cite{Boothby1987, Cawthorne1967, Russell1978, Mansur1978, Lee1992}, which leads to unacceptable embrittlement and loss of strength~\cite{Garner2012}.
G-phase grows in austenitic steel due to radiation induced segregation of Ni and Si~\cite{Maziasz1989a}.
However, the nucleation mechanism of G-phase from Ni-Si supersaturated regions is not well understood.
Moreover, the orientation relationship of the phase in austenite has not been established but is critical for understanding its detrimental effects.
The orientation relationship and particle morphology are prerequisite to determining the precipitate/matrix interfacial structure, which is ultimately responsible for the interaction with irradiation-induced point defects.

The cube-on-cube orientation relationship ~\cite{WILLIAMS1979} has been proposed to exist between G-phase and austenite, but most authors report there is no preferential orientation relationship~\cite{Lee1980, Maziasz1989a}.
Recent electron diffraction work has shown irradiation induced G-phase may have a preferential but complicated orientation relationship with austenite~\cite{Renault-Laborne2016, RenaultLaborne2014}; the Kurdjumov-Sachs (KS) orientation relationship ($(111)_\gamma || (1\bar{1}0)_\text{G}$, $[1\bar{1}0]_\gamma || [111]_\text{G}$) was suggested.
Orientation relationships with many variants like KS, Nishiyama-Wassermann ($(111)_\gamma || (1\bar{1}0)_\alpha$, $[11\bar{2}]_\gamma || [110]_\alpha$), Pitsch ($(100)_\gamma || (110)_\alpha$, $[011]_\gamma || [1\bar{1}1]_\alpha$) or Bain ($(100)_\gamma || (100)_\alpha$, $[010]_\gamma || [011]_\alpha$) are typically associated with the transformation of face-centered cubic (FCC) austenite ($\gamma$) to body-centered cubic (BCC) ferrite or martensite ($\alpha$).
Since G-phase forms with the cube-on-cube orientation relationship in ferrite, it is plausible to assume that G-phase may form in austenite analogously to ferrite~\cite{mateo1997characterization}.

The difficulty in establishing the orientation relationship of nano-sized precipitates with complex crystal structures using classical methods like selected area electron diffraction (SAED) is that diffraction patterns tend to contain many closely spaced reflections from multiple particles.
If the orientation relationships are non-trivial with multiple variants, indexing these patterns is extremely challenging, and it is impossible to tilt individual particles to different zone axis orientations.
An additional layer of complexity in this specific case is that irradiated austenite typically contains other precipitated phases with complex crystal structures and similar lattice parameter as G-phase, such as M$_{23}$C$_{6}$; these phases are often confused or misidentified.

In this paper, we attempted to elucidate the orientation relationship of nano-sized (10--20~nm) G-phase precipitates in ion-irradiated austenite using scanning nano-beam electron diffraction (NBED) inside a transmission electron microscope (TEM) utilizing a fast pixelated electron detector.
Scanning nano-beam diffraction is a technique whereby an electron probe with small convergence angle (typically $<1$~mrad) is scanned across the sample, and a diffraction pattern is collected at each scan position.
If the distance between the particles is large relative to the thickness of the specimen, then diffraction patterns will contain signal from individual particles and the matrix only.
By comparing the diffraction patterns to a library of simulated diffraction patterns of the crystal in different orientations (templates), local orientations in each scan position can be derived~\cite{Rauch2008, Rauch2010}.
The advantages of this technique are that no specific sample orientation is required before data collection, and crystallographic information from diffraction patterns can be directly correlated to real space coordinates.

One challenge of this method is that diffraction patterns contain both (weak) signal from the small precipitates and (strong) signal from the matrix phase above and below the particles.
Since some matrix reflections can correspond to some precipitate reflections and since they tend to be more intense than the precipitate reflections, this can confuse the template matching algorithm.
Recently it was demonstrated that this problem can be overcome by using a two stage indexation method~\cite{Valery2017, Rauch2019}.
In this method, the patterns are first indexed with the matrix templates, masked with the matrix template solution, and subsequently indexed with the precipitate phase templates.
Our initial trials showed that the conventional method of NBED data collection utilizing the ASTAR system (NanoMegas) was insufficiently sensitive to detect the faint G-phase reflections.
Reliable indexing of the diffraction patterns of small precipitates with complex crystal structure requires electron detection with the highest possible resolution and dynamic range.
Hence, diffraction pattern data was collected with the fast (up to 384 frames/s) TemCAM-XF416 pixelated CMOS detector (TVIPS).
A comparison of data collected with the ASTAR and TVIPS systems is given in supplementary materials.
The microscope used was a JEOL 2200FS TEM, equipped with a Schottky field emission gun and operating at 200~kV.
In addition to NBED, the morphology of the precipitates was investigated using high resolution high angle annular dark field (HR-HAADF) scanning TEM (STEM) using a probe corrected Titan microscope (Thermo Fisher Scientific) operating at 300 kV.

\begin{figure}
    \centering
    \includegraphics[width=0.8\linewidth]{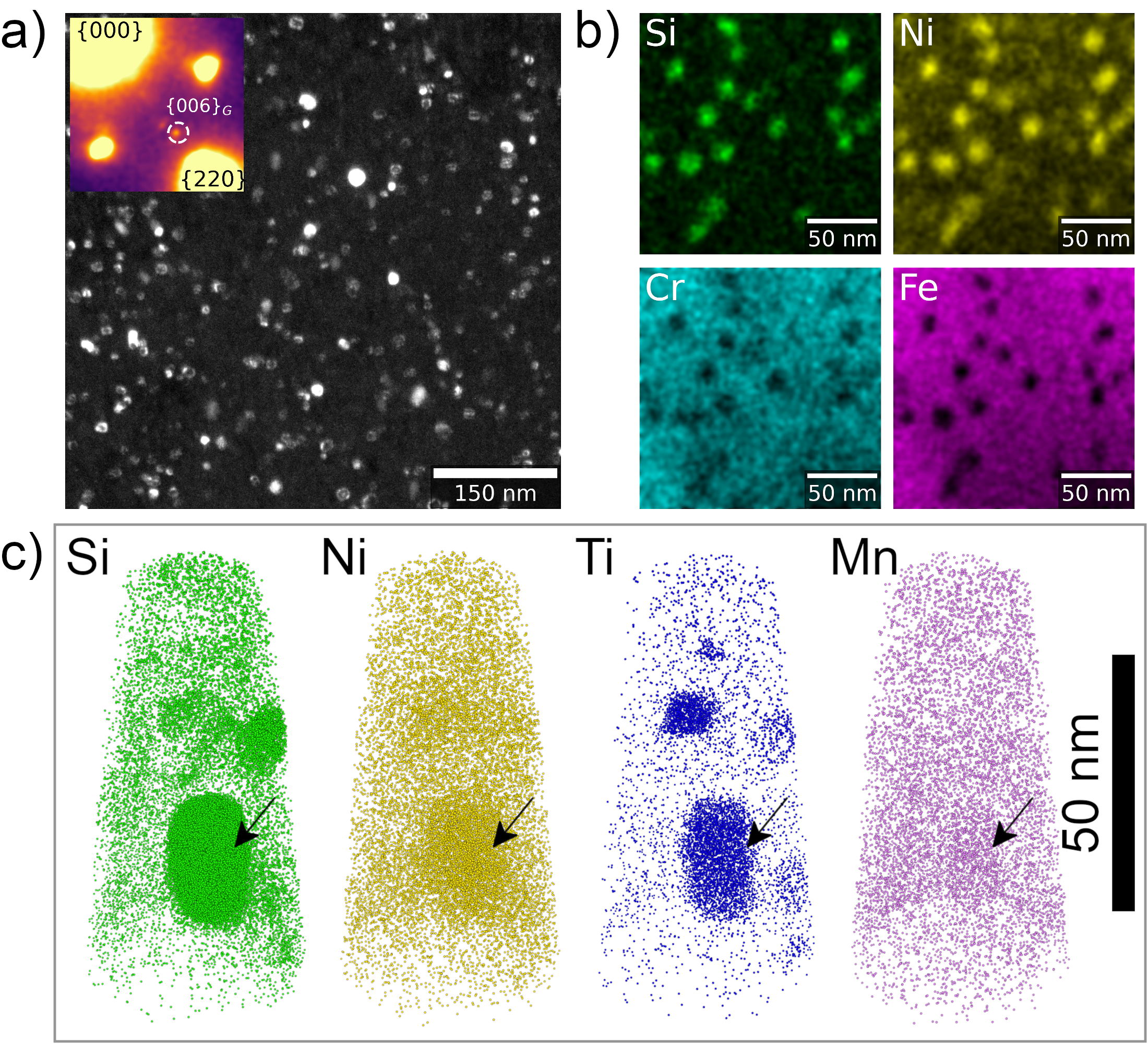}
    \caption{The presence of G-phase precipitates of 10-20~nm in size detected using three different methods. a) A TEM DF image by using a \{006\} G-phase reflection. Due to the different orientation variants, some of the precipitates are not visible in this image. b) STEM-EDX showing particles rich in Si and Ni and depleted in Cr and Fe. c) The same is observed in APT and also shows that the particles are enriched in Ti and Mn. The Fe and Cr ion maps are not shown.}
    \label{fig:gphasedemo}
\end{figure}

The material used in the study was an austenitic stainless steel in the 15--15Ti family with composition 15~wt\% Ni, 15~wt\% Cr, 1.8~wt\% Mn, 1.2~wt\% Mo, 0.5~wt\% Ti, 0.6~wt\% Si, 0.1~wt\% C, bal. Fe~\cite{Delville2014}, with a grain size of 10--15~$\mu$m.
The material was irradiated with 4.5~MeV Fe$^{2+}$ ions up to 40~dpa at 600~$^\circ$C. 
Additional details about the material and its characterization were published elsewhere~\cite{Cautaerts2018, Cautaerts2018a, Cautaerts2020}.
Figure~\ref{fig:gphasedemo} shows the G-phase precipitates as imaged in a) TEM dark field (DF), b) using scanning TEM (STEM) energy dispersive X-ray spectroscopy (EDX), and c) atom probe tomography (APT).
The precipitates were 10--20~nm in diameter, enriched in Si, Ni, Ti and Mn, and depleted in all other alloying elements.
No other irradiation induced phases were found in this steel, making this material ideally suited to study G-phase.
Electron transparent samples for TEM investigations were prepared via focused ion beam cross-section lift-out in a SCIOS 2 HiVac (Thermo Fisher Scientific) using a final cleaning at 2~kV.

The NBED data was collected using a camera length of 80~cm and a convergence angle of approximately 0.5~mrad.
A map of $260 \times 200$ scan positions was collected with a step size between 1-2~nm.
The region of interest was tilted slightly away from a $\langle 110 \rangle$ zone axis so that only few matrix reflections would be strongly excited.
The raw data from the camera was converted to the .BLO file format compatible with the ASTAR software (NanoMegas) using a custom tool~\cite{din14970-2020-4068502} and analyzed with methods and tools described in refs.~\cite{Valery2017, Rauch2019}.
This conversion involves some information loss, since the TVIPS camera collects images at 16-bit pixel depth, but the BLO format only supports 8-bit images.

\begin{figure}
    \centering
    \includegraphics[width=0.6\linewidth]{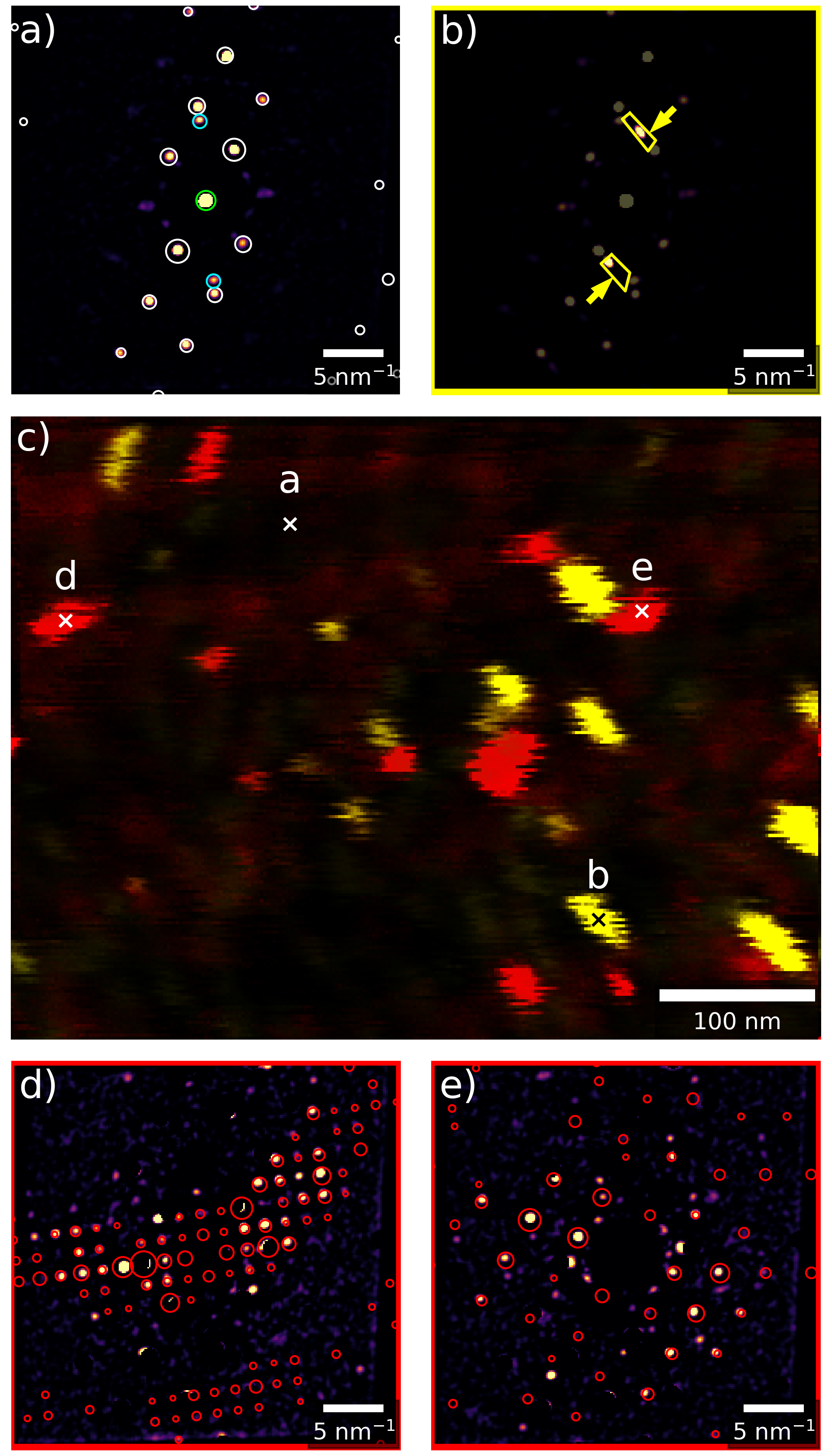}
    \caption{Illustration of the indexing procedure and exploration of the dataset.\ a) A diffraction pattern from the matrix with the best fit template superimposed as white circles. Additional anomalous spots and the direct beam are also circled in cyan and green respectively.\ b) A diffraction pattern showing additional diffraction spots originating from edge on Frank loops. The mask to select these features is superimposed.\ c) A composite VDF image. The yellow contribution represents the Frank loops as isolated by the mask shown in b). The red contribution represents the G-phase VDF obtained by masking the matrix and direct beam and Frank loops.\  d) and e) show diffraction patterns from locations containing G-phase particles. The best fit G-phase template is superimposed. Note that the background from inelastic scattering was removed from the diffraction patterns in a), b), d) and e) using a difference of Gaussians process.
}
    \label{fig:exploratory}
\end{figure}

The two stage process for indexing the G-phase is illustrated in Figure~\ref{fig:exploratory} (prepared with Hyperspy~\cite{francisco-de-la-pena-2020-3973513, de2017electron} and Pyxem~\cite{duncan-n-johnstone-2020-3976823}).
First, the austenite matrix is indexed in each pattern. 
Figure~\ref{fig:exploratory} a) shows a diffraction pattern from the matrix and the best fit austenite template as white circles.
The direct beam is circled with green.
Two additional diffraction spots were detected near the \{220\} reflections and circled with cyan.
These were found in all diffraction patterns and did not correspond to any distinguishable features in DF images; possibly they originate from surface oxide or small scale point defect clusters.
After the first step, the matrix contribution is removed with a mask created from the best fit template.
Signal that remains in the diffraction patterns originates from the G-phase particles and Frank loop defects (which also form under irradiation).
The Frank loops can be isolated by masking everything except the strip between the \{220\} and \{111\} matrix reflections as shown in Figure~\ref{fig:exploratory} b). 
The virtual dark field image (VDF) from the Frank loops is shown in yellow in Figure~\ref{fig:exploratory} c); since they grow on \{111\} planes they are elongated perpendicular to this direction.
A dark field TEM image demonstrating the presence of Frank loops in the irradiated layer can be found in supplementary materials.
Finally in the second stage, the G-phase is indexed.
The G-phase VDF is shown in red in Figure~\ref{fig:exploratory} c).
Two patterns from G-phase particles with the matrix contribution subtracted are shown in Figure~\ref{fig:exploratory} d) and e).
The best fit (highest matching index) G-phase templates are superimposed on the images.
Note that all features in the VDF image show jerky distortions in the x direction due to scan noise in the scanning system at high magnification, but since each diffraction pattern image is independent this does not impact the indexing process.

\begin{figure}
    \centering
    \includegraphics[width=\linewidth]{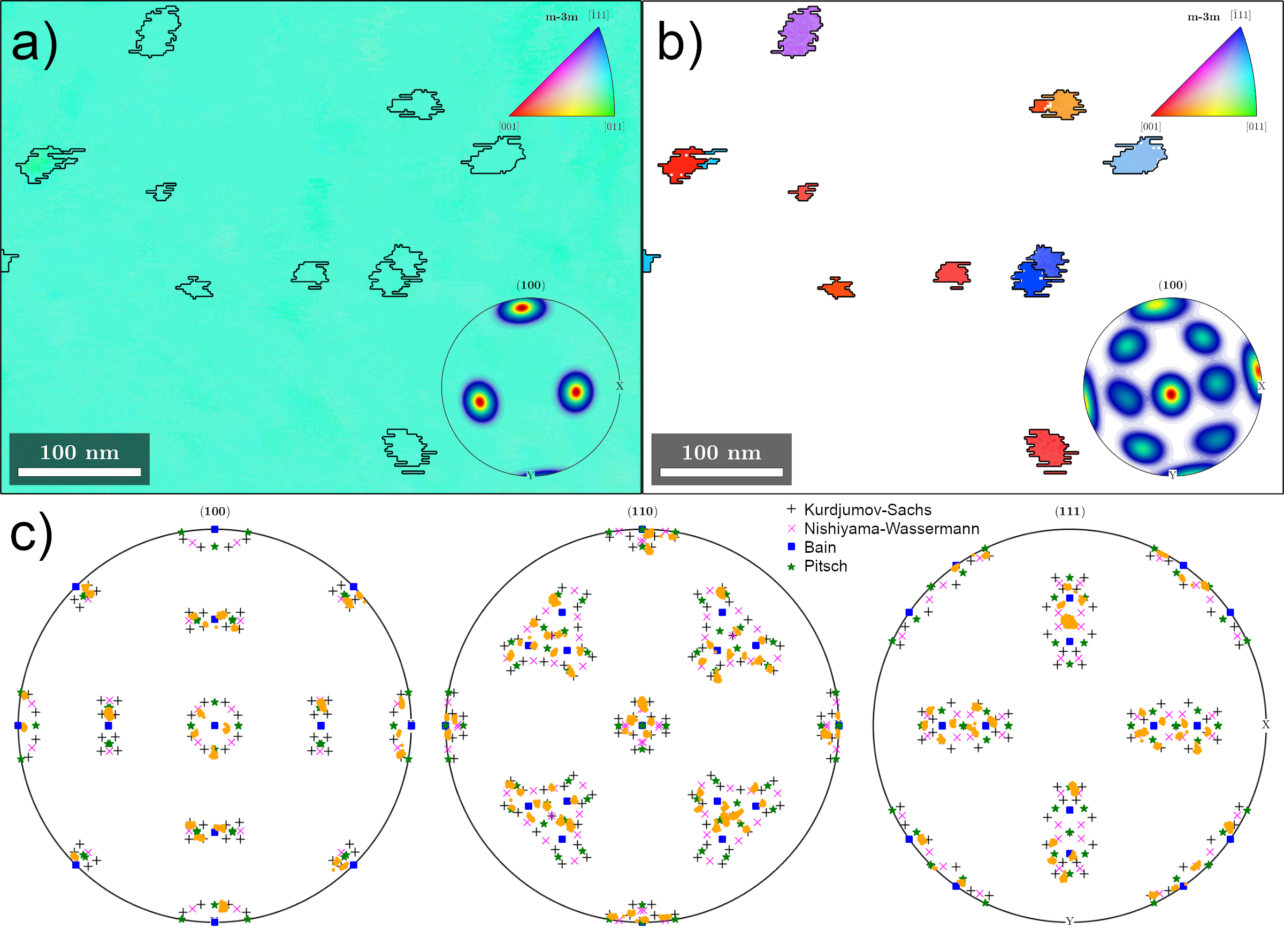}
    \caption{The orientation results analyzed with MTEX. Shown are the IPF-Z plots of (a) the austenite and (b) the G-phase. In each case, the \{100\} pole figures are shown in the bottom right corner and the IPF color legend in the upper right corner. Pixels with too low matching index were filtered out of the G-phase map to ensure only the regions containing particles are considered. (c) The misorientation between G-phase and austenite is plotted in \{100\}, \{110\} and \{111\} pole figures. Each orientation pixel is represented as a small orange dot, which are clustered into point clouds representing individual particles. For comparison, the variants of the most common FCC to BCC orientation relationships are also included in the plot.}
    \label{fig:mtexanalysis}
\end{figure}

The orientation maps were processed with MTEX v5.4.0~\cite{Bachmann2010}.
Figure~\ref{fig:mtexanalysis} summarizes the orientation results derived from the template matching procedure.
Figure~\ref{fig:mtexanalysis} a) shows the austenite inverse pole figure (IPF) map and in the bottom-right corner the (100) pole figure; both confirm that the matrix was oriented a few degrees away from a $\langle 110 \rangle$ zone axis.
The outlines of detected G-phase particles are superimposed on the plot.
The misorientation within the austenite grain spanned a range of about $\pm5^\circ$ relative to the mean orientation.
Figure~\ref{fig:mtexanalysis} b) shows a filtered G-phase inverse pole figure map and the (100) pole figure in the bottom right corner.
Pixels with a low correlation match index were excluded.
The map corresponds closely to the red features in the VDF image in Figure~\ref{fig:exploratory} c).
The data shows that some particles are oriented closely to a $\langle 100 \rangle$ zone, while others are oriented close to a $\langle 111 \rangle$ zone.
From the pole figure it appears that some $\langle 100 \rangle$ directions of the G-phase align closely with the $\langle 100 \rangle$ directions of the matrix as well as the $\langle 110 \rangle$ directions.
Figure~\ref{fig:mtexanalysis} c) represents the relative orientation between the G-phase and the matrix in each pixel.
The pole figures should be interpreted as if material in each pixel were rotated such that the austenite is aligned with the image coordinate system.
Each orange dot represents a G-phase pole from an indexed pixel; individual particles are thus represented by small orange point clouds.
Also shown on the plot for comparison are all the variants of the most common idealized orientation relationships between FCC and BCC phases.
These poles are to be interpreted as the BCC poles if the FCC basis were aligned with the coordinate system of the plot.
The experimental data points all fall within the range of these ideal orientation relationships, which clearly demonstrates that G-phase forms with orientation relationships characteristic of the FCC to BCC transformation.

Hence, the results suggest there is an analogy between the way G-phase forms in austenite and the way ferrite or martensite do.
Despite the space groups of G-phase and austenite being the same (Fm$\bar{3}$m), the G-phase lattice most closely resembles a superstructure of ferrite; only minor atomic shuffles are required to transform a $4 \times 4 \times 4$ supercell of BCC ferrite into the G-phase unit cell~\cite{mateo1997characterization}.
If the G-phase structure is similar to a BCC supercell, this implies the formation of the phase must be associated with a similar kind of tetragonal distortion and rotation of the FCC lattice as when austenite transforms to ferrite, which explains the similarity between the orientation relationships.
For all but the smallest particles, these strains must be accommodated by interface dislocations or twinning in the precipitating phase, which to a large extent determine the precipitate morphology and habit plane~\cite{dahmen1982orientation, Weatherly1994}.

\begin{figure}
    \centering
    \includegraphics[width=\linewidth]{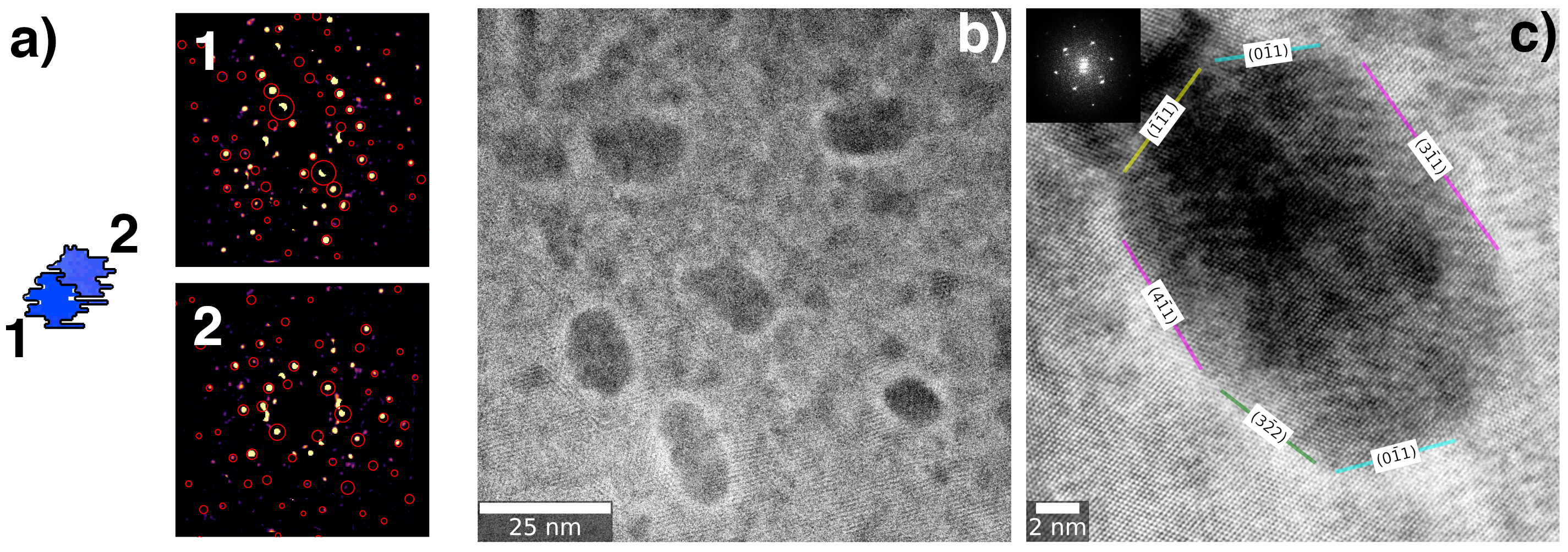}
    \caption{A more detailed view of the G-phase particles and their morphology. a) A precipitate seemingly composed of two grains. Diffraction patterns of both grains are shown on the right. The misorientation between the grains corresponds to a rotation of 52$^\circ$ over an axis close to the [111] direction in grain 1. b) A STEM-HAADF overview image showing a collection of G-phase particles with nearly round or elongated morphology. c) A HR-HAADF image of a large intragranular G-phase particle with the matrix in the $[011]$ zone orientation. The approximate projected interface planes are marked and labeled with miller indices in the matrix frame of reference. The fast Fourier transform of the matrix is shown in the inset.}
    \label{fig:twin_morphology}
\end{figure}

Some evidence of twinning inside G-phase precipitates was found, one example shown in Figure~\ref{fig:twin_morphology} a).
The two parts of the particle deviate 52$^\circ$ over a tilt axis close to the $[111]_\text{G}$ direction (nearly parallel to the optical axis) which resembles a $\Sigma 3$ twin of a 60$^\circ$ tilt over a $\langle 111 \rangle$ tilt axis.
The grains also appear to be joined on a plane parallel to \{111\}$_\gamma$ since the interface is parallel to the edge-on Frank loops in Figure~\ref{fig:exploratory}.
The deviation from the ideal twin rotation likely originates from a cumulation of errors in the indexing procedure; this could be the result of the relative weakness of the G-phase reflections, high concentration of defects and localized strains, the substantial image processing (including masking) prior to indexing the G-phase, and dynamic diffraction effects that are not accounted for in the template matching procedure.

It may be possible to further improve the angular resolution by optimizing the template matching routine to use mainly distant reflections, which are more sensitive to small changes in orientation~\cite{jeong2021automated}.
This could be coupled with more advanced acquisition strategies, such as acquiring multiple maps at different camera lengths optimized for each phase. 
In the dataset described here, the camera length is large for the matrix phase resulting in few matrix reflections in each pattern; this may increase uncertainty in determining the matrix orientation.
Furthermore, a more quantitative estimate of the angular resolution might be obtained via automated parameter studies; an enabling solution is being developed within the Pyxem library~\cite{duncan-n-johnstone-2020-3976823}.

FCC to BCC transformations tend to be associated with irrational interface planes containing an invariant line along an irrational lattice vector which can be predicted from lattice parameter ratios~\cite{dahmen1982orientation}.
If it is assumed that the G-phase unit cell represents $4 \times 4 \times 4$ BCC unit cells and the ratio $a_{fcc}/a_{bcc} = 3.6 / (11.5/4) = 1.25$, plate-shaped morphologies with near KS relationship and an invariant line close to $\langle 1\ 7\ 8  \rangle$ in the FCC lattice (or $\langle 1 1 1 \rangle$ in the BCC lattice) are expected.
Precipitates imaged with STEM-HAADF tended to be nearly round or slightly elongated in projection as shown in Figure~\ref{fig:twin_morphology} b), which is consistent with a plate-shaped morphology. 
A higher magnification image of one particle is shown in Figure~\ref{fig:twin_morphology} c).
The projected interface is demarcated by colored lines and labeled with Miller indices of the projected planes with respect to the austenite lattice.
The particles do not show clearly resolved low index facets, which is consistent with the theory of an irrational habit plane containing an invariant line.
The shape of the particle in Figure~\ref{fig:twin_morphology} c) would be consistent with a plate-shaped particle with invariant line close to $[\bar{1}10]_\gamma$ and habit plane between $(223)_\gamma$ and $(335)_\gamma$, inclined at an angle of approximately $30^\circ$ with the viewing direction.
In this scenario, a circular platelet would have an aspect ratio of 0.86, which is quite close to the observed value of 0.7 in the particle.
Unambiguous identification of the habit plane and the invariant line from the projections in these images may not be possible, but the findings seem consistent with theory~\cite{dahmen1982orientation} and observations in other FCC-BCC systems~\cite{Weatherly1994}.
A more precise confirmation of the orientation relationship of G-phase would require very thin regions (where the precipitate and matrix do not overlap) to be investigated along a $\langle 111 \rangle _\gamma$ zone axis with HR-(S)TEM -- approximately 1/4 of the particles should be oriented along a $\langle 111 \rangle _\text{G}$ zone if the particles follow near NW or KS relationships.

In summary, nano-sized irradiation induced G-phase precipitates were found to follow orientation relationships in austenite characteristic of the FCC to BCC transformation through NBED using a pixelated detector.
Only by collecting the data using a CMOS camera with high signal-to-noise and high dynamic range could the faint G-phase reflections be adequately resolved to index them with the template matching procedure.
If invariant line theory can be extended to this system, the KS orientation relationship should be preferred and the precipitates should have a plate-like morphology, which was confirmed by HR-HAADF imaging.
This information could be supplied to density functional theory or molecular dynamics simulations to investigate the energies and the interaction with radiation induced point defects of the precipitate/matrix interfaces, which might ultimately explain the relationship between this phase, void swelling and property degradation.
In the future, resolving individual orientation relationship variants directly with 4D-STEM may be enabled with finetuning the template matching procedure and data acquisition strategies.
This case study also shows the promise of this technique for investigating complex nano-sized precipitates in other material systems, for example in magnesium, aluminium, or titanium alloys~\cite{nie2012precipitation, ma2021precipitation, lendvai1996precipitation}.
Investigating the early stages of precipitation or quantifying orientation relationship variants once they can be resolved may be interesting application domains.


\section*{Acknowledgements}
We acknowledge funding through BiGmax (https://www.bigmax.mpg.de/), the Max Planck research network on big-data driven materials science.
This work was in part supported by the U.S. Department of Energy, Office of Nuclear Energy under DOE Idaho Operations Office Contract DE-AC07-051D14517 as part of a Nuclear Science User Facilities experiment.
We want to acknowledge SCK-CEN and the MYRRHA project for supplying the material, the staff at MIBL for assisting with the ion irradiations, and Dr. X. Zhou for the valuable discussion.

\bibliographystyle{model1a-num-names}
\bibliography{references}

\begin{thebibliography}{38}
\expandafter\ifx\csname natexlab\endcsname\relax\def\natexlab#1{#1}\fi
\providecommand{\url}[1]{\texttt{#1}}
\providecommand{\href}[2]{#2}
\providecommand{\path}[1]{#1}
\providecommand{\DOIprefix}{doi:}
\providecommand{\ArXivprefix}{arXiv:}
\providecommand{\URLprefix}{URL: }
\providecommand{\Pubmedprefix}{pmid:}
\providecommand{\doi}[1]{\href{http://dx.doi.org/#1}{\path{#1}}}
\providecommand{\Pubmed}[1]{\href{pmid:#1}{\path{#1}}}
\providecommand{\bibinfo}[2]{#2}
\ifx\xfnm\relax \def\xfnm[#1]{\unskip,\space#1}\fi
\bibitem[{Yan et~al.(2009)Yan, Grytsiv, Rogl, Pomjakushin, and
  Xue}]{yan2009crystal}
\bibinfo{author}{X.~Yan}, \bibinfo{author}{A.~Grytsiv},
  \bibinfo{author}{P.~Rogl}, \bibinfo{author}{V.~Pomjakushin},
  \bibinfo{author}{X.~Xue}, \bibinfo{journal}{Journal of alloys and compounds}
  \bibinfo{volume}{469} (\bibinfo{year}{2009}) \bibinfo{pages}{152--155}.
\bibitem[{Auger et~al.(1990)Auger, Danoix, Menand, Bonnet, Bourgoin, and
  Guttmann}]{auger1990atom}
\bibinfo{author}{P.~Auger}, \bibinfo{author}{F.~Danoix},
  \bibinfo{author}{A.~Menand}, \bibinfo{author}{S.~Bonnet},
  \bibinfo{author}{J.~Bourgoin}, \bibinfo{author}{M.~Guttmann},
  \bibinfo{journal}{Materials Science and Technology} \bibinfo{volume}{6}
  (\bibinfo{year}{1990}) \bibinfo{pages}{301--313}.
\bibitem[{Mateo et~al.(1997)Mateo, Llanes, Anglada, Redjaimia, and
  Metauer}]{mateo1997characterization}
\bibinfo{author}{A.~Mateo}, \bibinfo{author}{L.~Llanes},
  \bibinfo{author}{M.~Anglada}, \bibinfo{author}{A.~Redjaimia},
  \bibinfo{author}{G.~Metauer}, \bibinfo{journal}{Journal of Materials Science}
  \bibinfo{volume}{32} (\bibinfo{year}{1997}) \bibinfo{pages}{4533--4540}.
\bibitem[{Spiegel(1963)}]{spiegel1963ternary}
\bibinfo{author}{F.~Spiegel}, \bibinfo{journal}{Trans. Met. Soc. AIME}
  \bibinfo{volume}{27} (\bibinfo{year}{1963}) \bibinfo{pages}{575}.
\bibitem[{Shuro et~al.(2012)Shuro, Kuo, Sasaki, Hono, Todaka, and
  Umemoto}]{Shuro2012}
\bibinfo{author}{I.~Shuro}, \bibinfo{author}{H.~Kuo},
  \bibinfo{author}{T.~Sasaki}, \bibinfo{author}{K.~Hono},
  \bibinfo{author}{Y.~Todaka}, \bibinfo{author}{M.~Umemoto},
  \bibinfo{journal}{Materials Science and Engineering A} \bibinfo{volume}{552}
  (\bibinfo{year}{2012}) \bibinfo{pages}{194--198}.
  \DOIprefix\doi{10.1016/j.msea.2012.05.030}.
\bibitem[{Shiao et~al.(1994)Shiao, Tsai, Kai, and Huang}]{shiao1994aging}
\bibinfo{author}{J.~Shiao}, \bibinfo{author}{C.~Tsai},
  \bibinfo{author}{J.~Kai}, \bibinfo{author}{J.~Huang},
  \bibinfo{journal}{Journal of Nuclear Materials} \bibinfo{volume}{217}
  (\bibinfo{year}{1994}) \bibinfo{pages}{269--278}.
\bibitem[{Sun et~al.(2017)Sun, Marceau, Styles, Barbier, and
  Hutchinson}]{sun2017g}
\bibinfo{author}{W.~Sun}, \bibinfo{author}{R.~Marceau},
  \bibinfo{author}{M.~Styles}, \bibinfo{author}{D.~Barbier},
  \bibinfo{author}{C.~Hutchinson}, \bibinfo{journal}{Acta Materialia}
  \bibinfo{volume}{130} (\bibinfo{year}{2017}) \bibinfo{pages}{28--46}.
\bibitem[{Lee et~al.(1980)Lee, Maziasz, and Rowcliffe}]{Lee1980}
\bibinfo{author}{E.~Lee}, \bibinfo{author}{P.~Maziasz},
  \bibinfo{author}{A.~Rowcliffe}, \bibinfo{title}{{The Structure and
  Composition of Phases Occurring in Austenitic Stainless Steels in Thermal and
  Irradiation Environments}}, \bibinfo{type}{Technical Report}, Oak Ridge
  National Lab., \bibinfo{year}{1980}. \URLprefix
  \url{https://www.osti.gov/biblio/5889791}.
\bibitem[{Boothby and Williams(1988)}]{Boothby1987}
\bibinfo{author}{R.~M. Boothby}, \bibinfo{author}{T.~M. Williams},
  \bibinfo{journal}{Journal of Nuclear Materials} \bibinfo{volume}{152}
  (\bibinfo{year}{1988}) \bibinfo{pages}{123--138}.
\bibitem[{Lee et~al.(1983)Lee, Packan, and Mansur}]{Lee1983}
\bibinfo{author}{E.~H. Lee}, \bibinfo{author}{N.~H. Packan},
  \bibinfo{author}{L.~K. Mansur}, \bibinfo{journal}{Journal of Nuclear
  Materials} \bibinfo{volume}{117} (\bibinfo{year}{1983})
  \bibinfo{pages}{123--133}. \DOIprefix\doi{10.1016/0022-3115(83)90018-1}.
\bibitem[{Cawthorne and Fulton(1967)}]{Cawthorne1967}
\bibinfo{author}{C.~Cawthorne}, \bibinfo{author}{E.~J. Fulton},
  \bibinfo{journal}{Nature} \bibinfo{volume}{216} (\bibinfo{year}{1967})
  \bibinfo{pages}{575--576}. \DOIprefix\doi{10.1038/216575a0}.
\bibitem[{Russell(1978)}]{Russell1978}
\bibinfo{author}{K.~Russell}, \bibinfo{journal}{Acta Metallurgica}
  \bibinfo{volume}{26} (\bibinfo{year}{1978}) \bibinfo{pages}{1615--1630}.
  \DOIprefix\doi{10.1016/0001-6160(78)90071-8}.
\bibitem[{Mansur(1978)}]{Mansur1978}
\bibinfo{author}{L.~Mansur}, \bibinfo{journal}{Nuclear Technology}
  \bibinfo{volume}{40} (\bibinfo{year}{1978}) \bibinfo{pages}{5--34}.
  \DOIprefix\doi{10.13182/NT78-2}.
\bibitem[{Lee and Mansur(1992)}]{Lee1992}
\bibinfo{author}{E.~Lee}, \bibinfo{author}{L.~Mansur},
  \bibinfo{journal}{Metallurgical Transactions A} \bibinfo{volume}{23}
  (\bibinfo{year}{1992}) \bibinfo{pages}{1977--1986}.
  \DOIprefix\doi{10.1007/BF02647545}.
\bibitem[{Garner(2012)}]{Garner2012}
\bibinfo{author}{F.~Garner}, in: \bibinfo{editor}{R.~Konings} (Ed.),
  \bibinfo{booktitle}{Comprehensive Nuclear Materials},
  \bibinfo{publisher}{Elsevier}, \bibinfo{address}{New York},
  \bibinfo{year}{2012}, pp. \bibinfo{pages}{1--63}.
\bibitem[{Maziasz(1989)}]{Maziasz1989a}
\bibinfo{author}{P.~J. Maziasz}, \bibinfo{journal}{Journal of Nuclear
  Materials} \bibinfo{volume}{169} (\bibinfo{year}{1989})
  \bibinfo{pages}{95--115}. \DOIprefix\doi{10.1016/0022-3115(89)90525-4}.
\bibitem[{Williams et~al.(1979)Williams, Titchmarsh, and Arkell}]{WILLIAMS1979}
\bibinfo{author}{T.~Williams}, \bibinfo{author}{J.~Titchmarsh},
  \bibinfo{author}{D.~Arkell}, \bibinfo{journal}{Journal of Nuclear Materials}
  \bibinfo{volume}{82} (\bibinfo{year}{1979}) \bibinfo{pages}{199 -- 201}.
  \DOIprefix\doi{10.1016/0022-3115(79)90055-2}.
\bibitem[{Renault-Laborne et~al.(2016)Renault-Laborne, Garnier, Malaplate,
  Gavoille, Sefta, and Tanguy}]{Renault-Laborne2016}
\bibinfo{author}{A.~Renault-Laborne}, \bibinfo{author}{J.~Garnier},
  \bibinfo{author}{J.~Malaplate}, \bibinfo{author}{P.~Gavoille},
  \bibinfo{author}{F.~Sefta}, \bibinfo{author}{B.~Tanguy},
  \bibinfo{journal}{Journal of Nuclear Materials} \bibinfo{volume}{475}
  (\bibinfo{year}{2016}) \bibinfo{pages}{209--226}.
  \DOIprefix\doi{10.1016/j.jnucmat.2016.04.020}.
\bibitem[{Renault-Laborne et~al.(2014)Renault-Laborne, Malaplate, Pokor, and
  Tanguy}]{RenaultLaborne2014}
\bibinfo{author}{A.~Renault-Laborne}, \bibinfo{author}{J.~Malaplate},
  \bibinfo{author}{C.~Pokor}, \bibinfo{author}{B.~Tanguy},
  \bibinfo{journal}{Effects of Radiation on Nuclear Materials}
  (\bibinfo{year}{2014}) \bibinfo{pages}{1--24}.
  \DOIprefix\doi{10.1520/stp157220130096}.
\bibitem[{Rauch et~al.(2008)Rauch, V{\'{e}}ron, Portillo, Bultreys, Maniette,
  and Nicolopoulos}]{Rauch2008}
\bibinfo{author}{E.~Rauch}, \bibinfo{author}{M.~V{\'{e}}ron},
  \bibinfo{author}{J.~Portillo}, \bibinfo{author}{D.~Bultreys},
  \bibinfo{author}{Y.~Maniette}, \bibinfo{author}{S.~Nicolopoulos},
  \bibinfo{journal}{Microscopy and Analysis} \bibinfo{volume}{22}
  (\bibinfo{year}{2008}) \bibinfo{pages}{S5--S8}.
\bibitem[{Rauch et~al.(2010)Rauch, Portillo, Nicolopoulos, Bultreys, Rouvimov,
  and Moeck}]{Rauch2010}
\bibinfo{author}{E.~F. Rauch}, \bibinfo{author}{J.~Portillo},
  \bibinfo{author}{S.~Nicolopoulos}, \bibinfo{author}{D.~Bultreys},
  \bibinfo{author}{S.~Rouvimov}, \bibinfo{author}{P.~Moeck},
  \bibinfo{journal}{Zeitschrift fur Kristallographie} \bibinfo{volume}{225}
  (\bibinfo{year}{2010}) \bibinfo{pages}{103--109}.
  \DOIprefix\doi{10.1524/zkri.2010.1205}.
\bibitem[{Valery et~al.(2017)Valery, Rauch, Cl{\'{e}}ment, and
  Lorut}]{Valery2017}
\bibinfo{author}{A.~Valery}, \bibinfo{author}{E.~F. Rauch},
  \bibinfo{author}{L.~Cl{\'{e}}ment}, \bibinfo{author}{F.~Lorut},
  \bibinfo{journal}{Journal of Microscopy} \bibinfo{volume}{268}
  (\bibinfo{year}{2017}) \bibinfo{pages}{208--218}.
  \DOIprefix\doi{10.1111/jmi.12599}.
\bibitem[{Rauch and V{\'{e}}ron(2019)}]{Rauch2019}
\bibinfo{author}{E.~Rauch}, \bibinfo{author}{M.~V{\'{e}}ron},
  \bibinfo{journal}{Microscopy and Microanalysis} \bibinfo{volume}{25}
  (\bibinfo{year}{2019}) \bibinfo{pages}{1922--1923}.
  \DOIprefix\doi{10.1017/s1431927619010341}.
\bibitem[{Delville et~al.(2014)Delville, Stergar, and Verwerft}]{Delville2014}
\bibinfo{author}{R.~Delville}, \bibinfo{author}{E.~Stergar},
  \bibinfo{author}{M.~Verwerft}, in: \bibinfo{booktitle}{22nd International
  Conference on Nuclear Engineering ICONE22}, \bibinfo{publisher}{ASME},
  \bibinfo{address}{Prague, Czech Republic}, \bibinfo{year}{2014}.
\bibitem[{Cautaerts et~al.(2018{\natexlab{a}})Cautaerts, Delville, Stergar,
  Schryvers, and Verwerft}]{Cautaerts2018}
\bibinfo{author}{N.~Cautaerts}, \bibinfo{author}{R.~Delville},
  \bibinfo{author}{E.~Stergar}, \bibinfo{author}{D.~Schryvers},
  \bibinfo{author}{M.~Verwerft}, \bibinfo{journal}{Journal of Nuclear
  Materials} \bibinfo{volume}{507} (\bibinfo{year}{2018}{\natexlab{a}})
  \bibinfo{pages}{177--187}. \DOIprefix\doi{10.1016/j.jnucmat.2018.04.041}.
\bibitem[{Cautaerts et~al.(2018{\natexlab{b}})Cautaerts, Delville, Stergar,
  Schryvers, and Verwerft}]{Cautaerts2018a}
\bibinfo{author}{N.~Cautaerts}, \bibinfo{author}{R.~Delville},
  \bibinfo{author}{E.~Stergar}, \bibinfo{author}{D.~Schryvers},
  \bibinfo{author}{M.~Verwerft}, \bibinfo{journal}{Acta Materialia}
  \bibinfo{volume}{164} (\bibinfo{year}{2018}{\natexlab{b}})
  \bibinfo{pages}{90--98}. \DOIprefix\doi{10.1016/J.ACTAMAT.2018.10.018}.
\bibitem[{Cautaerts et~al.(2020)Cautaerts, Delville, Stergar, Pakarinen,
  Verwerft, Yang, Hofer, Schnitzer, Lamm, Felfer, and
  Schryvers}]{Cautaerts2020}
\bibinfo{author}{N.~Cautaerts}, \bibinfo{author}{R.~Delville},
  \bibinfo{author}{E.~Stergar}, \bibinfo{author}{J.~Pakarinen},
  \bibinfo{author}{M.~Verwerft}, \bibinfo{author}{Y.~Yang},
  \bibinfo{author}{C.~Hofer}, \bibinfo{author}{R.~Schnitzer},
  \bibinfo{author}{S.~Lamm}, \bibinfo{author}{P.~Felfer},
  \bibinfo{author}{D.~Schryvers}, \bibinfo{journal}{Acta Materialia}
  \bibinfo{volume}{197} (\bibinfo{year}{2020}) \bibinfo{pages}{184--197}.
  \DOIprefix\doi{10.1016/j.actamat.2020.07.022}.
\bibitem[{Cautaerts and Harrison(2020)}]{din14970-2020-4068502}
\bibinfo{author}{N.~Cautaerts}, \bibinfo{author}{P.~Harrison},
  \bibinfo{title}{din14970/tvipsconverter: tvipsconverter v0.1.3},
  \bibinfo{year}{2020}. \DOIprefix\doi{10.5281/zenodo.4288857}.
\bibitem[{de~la Pe{\~n}a et~al.(2020)de~la Pe{\~n}a, Prestat, Fauske, Burdet,
  Furnival, Jokubauskas, Nord, Ostasevicius, MacArthur, Johnstone, Sarahan,
  L{\"a}hnemann, Taillon, pquinn dls, Aarholt, Migunov, Eljarrat, Caron,
  Mazzucco, Martineau, Somnath, Poon, Walls, Slater, Tappy, Cautaerts, Winkler,
  Donval, and Myers}]{francisco-de-la-pena-2020-3973513}
\bibinfo{author}{F.~de~la Pe{\~n}a}, \bibinfo{author}{E.~Prestat},
  \bibinfo{author}{V.~T. Fauske}, \bibinfo{author}{P.~Burdet},
  \bibinfo{author}{T.~Furnival}, \bibinfo{author}{P.~Jokubauskas},
  \bibinfo{author}{M.~Nord}, \bibinfo{author}{T.~Ostasevicius},
  \bibinfo{author}{K.~E. MacArthur}, \bibinfo{author}{D.~N. Johnstone},
  \bibinfo{author}{M.~Sarahan}, \bibinfo{author}{J.~L{\"a}hnemann},
  \bibinfo{author}{J.~Taillon}, \bibinfo{author}{pquinn dls},
  \bibinfo{author}{T.~Aarholt}, \bibinfo{author}{V.~Migunov},
  \bibinfo{author}{A.~Eljarrat}, \bibinfo{author}{J.~Caron},
  \bibinfo{author}{S.~Mazzucco}, \bibinfo{author}{B.~Martineau},
  \bibinfo{author}{S.~Somnath}, \bibinfo{author}{T.~Poon},
  \bibinfo{author}{M.~Walls}, \bibinfo{author}{T.~Slater},
  \bibinfo{author}{N.~Tappy}, \bibinfo{author}{N.~Cautaerts},
  \bibinfo{author}{F.~Winkler}, \bibinfo{author}{G.~Donval},
  \bibinfo{author}{J.~C. Myers}, \bibinfo{title}{hyperspy/hyperspy: Release
  v1.6.1}, \bibinfo{year}{2020}. \DOIprefix\doi{10.5281/zenodo.4294676}.
\bibitem[{de~la Pe{\~n}a et~al.(2017)de~la Pe{\~n}a, Ostasevicius, Fauske,
  Burdet, Jokubauskas, Nord, Sarahan, Prestat, Johnstone, Taillon
  et~al.}]{de2017electron}
\bibinfo{author}{F.~de~la Pe{\~n}a}, \bibinfo{author}{T.~Ostasevicius},
  \bibinfo{author}{V.~T. Fauske}, \bibinfo{author}{P.~Burdet},
  \bibinfo{author}{P.~Jokubauskas}, \bibinfo{author}{M.~Nord},
  \bibinfo{author}{M.~Sarahan}, \bibinfo{author}{E.~Prestat},
  \bibinfo{author}{D.~N. Johnstone}, \bibinfo{author}{J.~Taillon}, et~al.,
  \bibinfo{journal}{Microscopy and Microanalysis} \bibinfo{volume}{23}
  (\bibinfo{year}{2017}) \bibinfo{pages}{214--215}.
\bibitem[{Johnstone et~al.(2021)Johnstone, Crout, Nord, Laulainen,
  H{\o}g{\aa}s, Opheim, Martineau, Francis, Bergh, Prestat, Smeets, Ross,
  Collins, Hjorth, Mohsen, Furnival, Jannis, Cautaerts, Jacobsen, Herzing,
  Poon, {\AA}nes, Morzy, Doherty, Iqbal, Ostasevicius, von Lany, and
  Tovey}]{duncan-n-johnstone-2020-3976823}
\bibinfo{author}{D.~N. Johnstone}, \bibinfo{author}{P.~Crout},
  \bibinfo{author}{M.~Nord}, \bibinfo{author}{J.~Laulainen},
  \bibinfo{author}{S.~H{\o}g{\aa}s}, \bibinfo{author}{E.~Opheim},
  \bibinfo{author}{B.~Martineau}, \bibinfo{author}{C.~Francis},
  \bibinfo{author}{T.~Bergh}, \bibinfo{author}{E.~Prestat},
  \bibinfo{author}{S.~Smeets}, \bibinfo{author}{A.~Ross},
  \bibinfo{author}{S.~Collins}, \bibinfo{author}{I.~Hjorth},
  \bibinfo{author}{Mohsen}, \bibinfo{author}{T.~Furnival},
  \bibinfo{author}{D.~Jannis}, \bibinfo{author}{N.~Cautaerts},
  \bibinfo{author}{E.~Jacobsen}, \bibinfo{author}{A.~Herzing},
  \bibinfo{author}{T.~Poon}, \bibinfo{author}{H.~W. {\AA}nes},
  \bibinfo{author}{J.~Morzy}, \bibinfo{author}{T.~Doherty},
  \bibinfo{author}{A.~Iqbal}, \bibinfo{author}{T.~Ostasevicius},
  \bibinfo{author}{M.~von Lany}, \bibinfo{author}{R.~Tovey},
  \bibinfo{title}{pyxem/pyxem: pyxem 0.13.0}, \bibinfo{year}{2021}.
  \DOIprefix\doi{10.5281/zenodo.4436723}.
\bibitem[{Bachmann et~al.(2010)Bachmann, Hielscher, and
  Schaeben}]{Bachmann2010}
\bibinfo{author}{F.~Bachmann}, \bibinfo{author}{R.~Hielscher},
  \bibinfo{author}{H.~Schaeben}, \bibinfo{journal}{Solid State Phenomena}
  \bibinfo{volume}{160} (\bibinfo{year}{2010}) \bibinfo{pages}{63--68}.
  \DOIprefix\doi{10.4028/www.scientific.net/SSP.160.63}.
\bibitem[{Dahmen(1982)}]{dahmen1982orientation}
\bibinfo{author}{U.~Dahmen}, \bibinfo{journal}{Acta Metallurgica}
  \bibinfo{volume}{30} (\bibinfo{year}{1982}) \bibinfo{pages}{63--73}.
\bibitem[{Weatherly and Zhang(1994)}]{Weatherly1994}
\bibinfo{author}{G.~C. Weatherly}, \bibinfo{author}{W.~Z. Zhang},
  \bibinfo{journal}{Metallurgical and Materials Transactions A}
  \bibinfo{volume}{25} (\bibinfo{year}{1994}) \bibinfo{pages}{1865--1874}.
  \DOIprefix\doi{10.1007/BF02649034}.
\bibitem[{Jeong et~al.(2021)Jeong, Cautaerts, Dehm, and
  Liebscher}]{jeong2021automated}
\bibinfo{author}{J.~Jeong}, \bibinfo{author}{N.~Cautaerts},
  \bibinfo{author}{G.~Dehm}, \bibinfo{author}{C.~H. Liebscher},
  \bibinfo{journal}{arXiv preprint arXiv:2102.09711}  (\bibinfo{year}{2021}).
\bibitem[{Nie(2012)}]{nie2012precipitation}
\bibinfo{author}{J.-F. Nie}, \bibinfo{journal}{Metallurgical and Materials
  Transactions A} \bibinfo{volume}{43} (\bibinfo{year}{2012})
  \bibinfo{pages}{3891--3939}.
\bibitem[{Ma et~al.(2021)Ma, Zheng, Dasari, Zhang, Fraser, and
  Banerjee}]{ma2021precipitation}
\bibinfo{author}{K.~Ma}, \bibinfo{author}{Y.~Zheng},
  \bibinfo{author}{S.~Dasari}, \bibinfo{author}{D.~Zhang},
  \bibinfo{author}{H.~L. Fraser}, \bibinfo{author}{R.~Banerjee},
  \bibinfo{journal}{MRS Bulletin}  (\bibinfo{year}{2021})
  \bibinfo{pages}{1--8}.
\bibitem[{Lendvai(1996)}]{lendvai1996precipitation}
\bibinfo{author}{J.~Lendvai}, in: \bibinfo{booktitle}{Materials Science Forum},
  volume \bibinfo{volume}{217}, \bibinfo{organization}{Trans Tech Publ}, pp.
  \bibinfo{pages}{43--56}.

\end{thebibliography}


\end{document}